\documentclass[twocolumn,showpacs,preprintnumbers,amsmath,amssymb,article]{revtex4}

\usepackage{graphicx}
\usepackage{dcolumn}
\usepackage{bm}

\begin{document}

\preprint{APS/123-QED}

\title{Susceptibilities and fluctuations in a Quark-Hadron System\\ with Dynamical Recombination}

\author{Stephane Haussler}
\author{Marcus Bleicher}%
\affiliation{%
Institut f\"ur Theoretische Physik, Johann Wolfgang Goethe-Universit\"at, Max-von-Laue-Str.~1, D-60438 Frankfurt am Main, Germany
}%
\author{Horst St\"ocker}%
\affiliation{%
Gesellschaft fur Schwerionenforschung (GSI), Planck-Str.~1, D-64291 Darmstadt, Germany
}%
\affiliation{%
Institut f\"ur Theoretische Physik (ITP), Johann Wolfgang Goethe-Universit\"at, Max-von-Laue-Str.~1, D-60438 Frankfurt am Main, Germany
}%
\affiliation{%
Frankfurt Institute for Advanced Studies (FIAS), Ruth-Moufang-Str.~1, D-60438 Frankfurt am Main, Germany\\
}%

\date{\today}

\begin{abstract}
Within the framework of the dynamical recombination approach implemented in the partonic/hadronic quark Molecular Dynamics (qMD) model, we investigate numerous QGP signals constructed from the correlations and fluctuations of conserved charges, namely charged particle ratio fluctuations, charge transfer fluctuations as well as various ratios of susceptibilities. We argue that more generally, the covariances and the variances of the conserved charges divided by the number of charged particles are a measure of the quark number susceptibilities and are thus sensitive to the phase of the system. Computations carried within samples of central qMD events from low AGS energies on ($E_{lab}=2$~AGeV) up to the highest RHIC energies available ($\sqrt{s_{NN}}=200$~GeV) show that the final state calculations are always compatible with the hadronic result. From analyses performed as a function of time with our sample of events at $\sqrt{s_{NN}}=200$~GeV, we find the recombination-like hadronization process the qMD system undergoes to be responsible for the vanishing of the early stage QGP correlations and fluctuations. These results might explain the compatibility of the measurements on charged particle ratio fluctuations with the hadronic expectations and leave no room for the experimental observation of any QGP fluctuation/correlation signals based upon the susceptibilities of the conserved charges.
\end{abstract}



\maketitle

\section{Introduction}

There exist a large consensus in the heavy-ion physics community that a plasma of deconfined quarks and gluons is formed during the first instant of the collision of heavy nuclei for sufficiently high incoming energies \cite{Adams:2005dq,Adcox:2004mh,Arsene:2004fa}. The investigation of the properties of the highly excited nuclear matter produced requires the use of physical quantities sensitive to the initial partonic state of the system which survive hadronization and the subsequent hadronic evolution. 

Numerous probes based on the eventwise fluctuations of physical quantities have been proposed to investigate the state and the properties of the fireball \cite{Gazdzicki:1992ri,Mrowczynski:1997kz,Bleicher:1998wd,Bleicher:1998wu,Stephanov:1998dy,Jeon:1999gr,Stephanov:1999zu,Mrowczynski:1999sf,Capella:1999uc,Mrowczynski:1999un,Bleicher:2000tr,Bleicher:2000ek,Jeon:2001ka,Sa:2001ma,Koch:2001zn,Hatta:2003wn,Ferreiro:2003dw,Mrowczynski:2004cg,Konchakovski:2005hq,Cunqueiro:2005hx,Torrieri:2005va,Armesto:2006bv,Gorenstein:2007ep,Begun:2006jf,Lungwitz:2007uc,Hauer:2007im,Konchakovski:2007ah,Asakawa:2000wh,Haussler:2006mq,Haussler:2007un,Jeon:2005yi,Jeon:2005kj,Shi:2005rc,Majumder:2005ai,Koch:2005vg}. Among them, charged particle ratio fluctuations, charge transfer fluctuations, baryon number-strangeness correlations and charge-strangeness correlations were prominently proposed to pin down the formation of the deconfined phase at RHIC \cite{Jeon:2000wg,Asakawa:2000wh,Jeon:2001ue,Zhang:2002dy,Pruneau:2002yf,Haussler:2006mq,Haussler:2007un,Jeon:2005yi,Jeon:2005kj,Shi:2005rc,Majumder:2005ai,Koch:2005vg,Haussler:2005ei}. 
It was pointed out that these quantities should reflect the properties of the system in the first instants of the reaction and survive the whole evolution: with a strong longitudinal flow, any conserved charge is frozen in a given rapidity window because the expansion is too fast for the initial partonic fluctuations to relax through the transport of charges in and out of a given rapidity slice. Thus, QGP fluctuations should not have the possibility to relax to their hadronic expectation values. It is clear that the size of the rapidity window for fluctuation studies must not be too wide in order to avoid global charge conservation which would lead to a vanishing signal, but also neither too small to avoid purely statistical fluctuations and the transport of charges in and out of the slice under consideration by hadronic rescattering. The generally accepted rapidity width  is of the order of $\Delta y = 0.5-1$ units in rapidity.

However, hadronization itself might bring the initial QGP correlations and fluctuations to values compatible with the hadronic expectations. How to calculate or describe the non-perturbative domain of QCD where the physics of hadronization sits remains today an open question. A possible mechanism consists of the recombination of quarks into hadrons \cite{Biro:1994mp,Zimanyi:1999py,Hwa:2003ic,Fries:2003vb,Greco:2003xt,Molnar:2003ff,Fries:2003kq,Hofmann:1999jy,Scherer:2005sr,Scherer:2001ap,Hofmann:1999jx}. On the experimental side, the most striking evidences for recombination are the number-of-constituent-quarks scaling of the elliptic flow $v_2$ \cite{Adams:2003am,Molnar:2003ff,Krieg:2007sx} and the large number of baryons over number of mesons ratio in the intermediate $p_t$ range measured in central Au+Au collisions at RHIC energies \cite{Fries:2003kq,Greco:2003xt,Hwa:2004ng}. In this paper, we use the  quark Molecular Dynamics (qMD) model \cite{Hofmann:1999jy,Scherer:2005sr,Scherer:2001ap,Hofmann:1999jx} where quarks coalesce locally in  coordinate and momentum space due to a confining inter-quark potential.

In previous studies \cite{Haussler:2006mq,Haussler:2007un} performed with qMD and in an earlier exploratory work \cite{Nonaka:2005vr}, it was argued that recombination leads to a complete vanishing of various initial state QGP fluctuation and correlation signals. We extend this idea further and show that dynamical recombination generally blurs QGP signals constructed from the correlations and fluctuations of conserved charges.

Besides the predicted impossibility to experimentally establish the existence of QGP fluctuations, our results are important for the fluctuation investigations planned for the search of the critical endpoint of the QCD phase diagram at the forthcoming FAIR facility at GSI.

The article is organized as follow: We first introduce the quark Molecular Dynamics (qMD) model and address the problem of entropy conservation at the partonic/hadronic transition which might occur in some recombination approaches. We detail the situation for our dynamical recombination procedure and argue that we respect the necessary increase of entropy with time. Within our sample of events generated with the qMD model, we then investigate the charged particle ratio fluctuations, the charge transfer fluctuations, the baryon number-strangeness correlation coefficient, the charge-strangeness correlation coefficient and discuss the influence of recombination-hadronization for all these observables. We also compare the result of the qMD with lattice data \cite{Gavai:2005yk,Majumder:2006nq} on various ratios of susceptibilities and to directly investigate the susceptibilities by performing event-by-event analyses. Finally, we discuss the reasons for the transition from the initial QGP fluctuations and correlations to their hadronic value.

\section{The ${\textrm q}$MD model}

The qMD model \cite{Hofmann:1999jy,Scherer:2005sr,Scherer:2001ap,Hofmann:1999jx}  employed here is a semi-classical molecular dynamics approach where quarks are treated as point-like particles carrying color charges and interact via a linear heavy quark potential. Initial conditions for the qMD are taken from the hadron-string transport model UrQMD \cite{Bleicher:1999xi,Bass:1998ca}: After the two incoming nuclei have passed through each other, (pre-) hadrons from the strings and fully formed hadrons from the UrQMD model are decomposed into quarks with current masses $m_u=m_d=5$~MeV and $m_s=150$~MeV. At the highest RHIC energy, this happens at a center of mass time of $t=0.15$~fm/c. It should be noted that the qualitative results of the present study are not restricted to any specific initial state. The UrQMD model is solely used to provide an exemplary initial state after the initial $q\overline q$ production has taken place. The quarks are then let to evolve and interact within the qMD via a linear potential $V(|\bold{r}_i-\bold{r}_j|)=\kappa |\bold{r}_i-\bold{r}_j|$, where $\kappa$ is the string tension and $\bold{r}_n$ is the position of particle $n$. Therefore the full Hamiltonian of the model reads:
\begin{equation} H= \sum_{i=1}^N \sqrt{ p_i^2 + m_i^2} + \frac{1}{2}
\sum_{i,j} C_{ij} V(|\bold{r}_i-\bold{r}_j|)\quad.
\label{eq:hamiltonian}
\end{equation}
where $N$ counts the number of particles in the system and the term $C_{ij}$ takes into account the color dependence of the interaction.

The quark--(anti-)quark interaction within this potential naturally leads to confinement through the binding of (anti-)quarks into color neutral clusters. New hadrons are formed from quarks whose momentum and positions are close to each others. Typical values for the relative momenta of the quarks in the two-particle rest frame at hadronization are $|p_{q}|=|p_{\bar q}|\leq 500$~MeV, the typical distance is below 1~fm, i.e. hadronization occurs locally into hadronic clusters of mesonic and baryonic type that resemble the Yo-Yo states of the LUND model. Most of the mass of the newly formed hadrons come from the potential energy between the quarks forming the considered cluster. We generally obtain clusters with large masses mapped to hadronic resonances and thus describe the transition from current quark masses to constituent quark masses of the order of 300~MeV after hadronization has occured. These clusters are then mapped to hadron states that are later allowed to decay in the further evolution of the system. The reader is referred to \cite{Hofmann:1999jy,Scherer:2005sr,Scherer:2001ap,Hofmann:1999jx} for a detailed discussion of the qMD model. 

\section{Entropy Conservation}

The question of entropy at the parton/hadron transition needs to be examined when quark recombination is used to describe hadronization. Entropy is proportional to the number of particles in a gas of massless partons/hadrons. Thus, when partons recombine to form new hadrons, the total number of particles decreases and with it entropy. This situation is not acceptable as it violates the second law of thermodynamics.

Contrary to many other recombination models, one can argue that qMD does not violate the entropy condition. In fact, entropy does not only depend on the number of particles but also on their masses as examined in \cite{Nonaka:2005vr}. When $m/T>3$, entropy can be simply approximated for a gas of massive particles through:
\begin{equation}
S=N (3.5+m/T) \quad ,
\label{eq:entropy}
\end{equation}
where $S$ stands for entropy, $N$ for the number of particles in the system, $m$ for their average mass and $T$ for the temperature at the transition.

\begin{figure}
\includegraphics[width=0.5\textwidth]{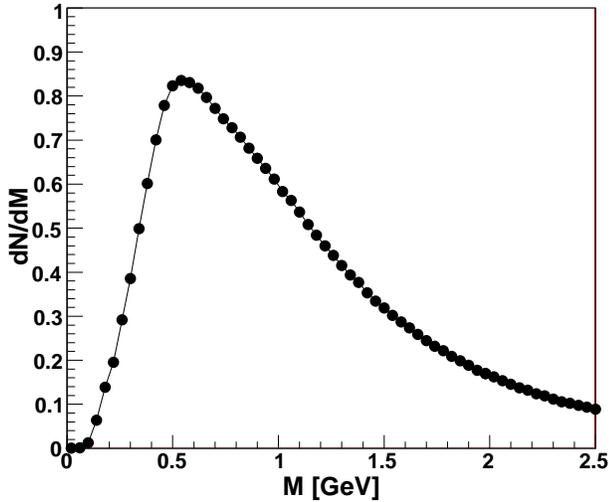}
\caption{\label{fig:ClusterMassDistribution} Distribution of the mass of the clusters obtained after recombination of the initial partons for central Au+Au qMD reactions at $\sqrt{s_{NN}}=200$~GeV.}
\end{figure}

Fig.~\ref{fig:ClusterMassDistribution} shows the distribution of the clusters masses at hadronization obtained from qMD central Au+Au simulations at $\sqrt{s}=200$~AGeV. The average mass after recombination of the quarks is large with $M_{cluster} \approx 730$~MeV.

To estimate entropy, let us consider a system of $N_{q+\bar{q}}$ quarks and anti-quarks that recombine into $N_{q+\bar{q}}/2$ hadrons where baryons are neglected for simplicity. The quarks being nearly massless, entropy in the partonic phase is equal to \cite{Nonaka:2005vr}:
\begin{equation}
S_{QGP}=4.2 N_{q+\bar{q}} \quad ,
\end{equation}
After recombination and using Eq.~\ref{eq:entropy} to estimate entropy in the hadronic phase:
\begin{equation}
S_{reco}=\frac{N_{q+\bar{q}}}{2}*(3.5+730/150)=4.2 N_{q+\bar{q}} \quad .
\end{equation}
where the qMD average cluster mass $m=730$~MeV and the qMD critical temperature $T=150$~MeV are used. Thus, $S_{reco} \approx S_{QGP}$ and entropy does not decrease in the process of recombining quarks into massive hadronic clusters. 

\begin{figure}
\includegraphics[width=0.5\textwidth]{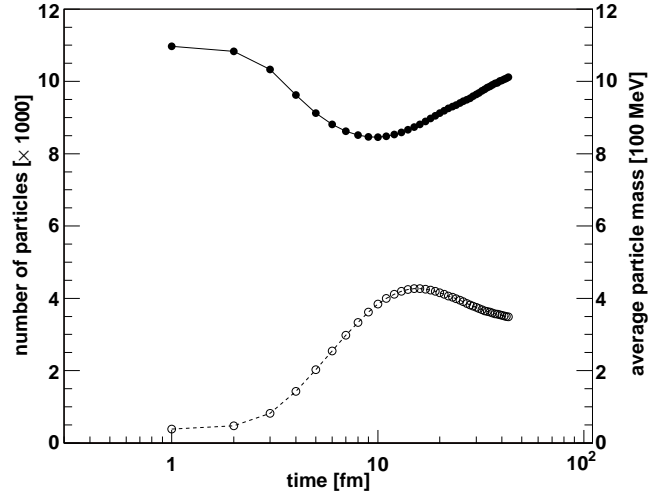}
\caption{\label{fig:particlesandmass} 
Total number of particles $N_{quarks}+N_{hadrons}$ (full circles) as a function of time and average mass as a function of time (open circles) for Au+Au reactions at $\sqrt{s_{NN}}=200$~GeV.}
\end{figure}

Entropy is further analyzed through the number of particles in the qMD model. Fig.~\ref{fig:particlesandmass} depicts the total number of particle (full circles) and the average particle mass (empty circles) as a function of time. The average particle mass peaks around a time $t \approx 11$ fm/c to an average mass of $\langle m \rangle \approx 450$~MeV. This is much less than the average cluster mass observed in Fig.~\ref{fig:ClusterMassDistribution} and is due to the decay of the excited clusters produced at hadronization. Further decrease of the average particle mass is related to the continuing decay of resonances when all particles are already confined into hadronic clusters. As can be seen from Fig.~\ref{fig:particlesandmass}, the total number of particles decreases at hadronization. However, resonance decay leads to a number of particles in the final state close to the initial number of quarks. The increase of entropy with time is then not violated, at least when comparing initial and final state.

Results from lattice QCD calculations \cite{Biro:2006sv} also indicate that recombination is compatible with entropy conservation at the transition.

\section{Fluctuations of conserved charges}

\begin{figure}
\includegraphics[width=0.5\textwidth]{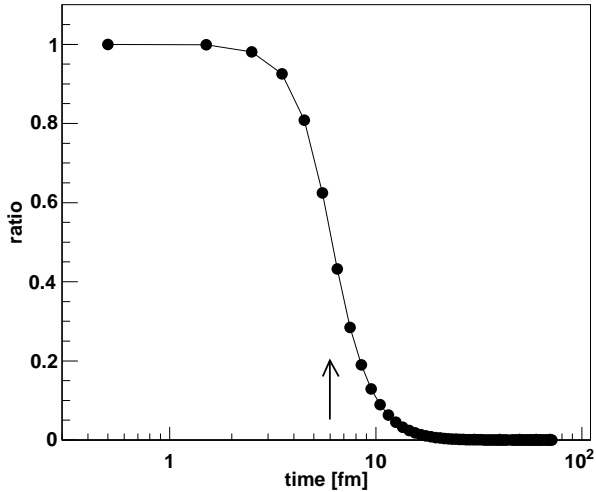}
\caption{\label{fig:ParticleNumber} 
Fraction of the number of quarks from the total number of 
particles $(N_q+N_{\bar{q}})/(N_q+N_{\bar{q}}+N_{H})$ as a function of time at midrapidity for central Au+Au qMD reactions at $\sqrt{s_{NN}}=200$~GeV. The arrow depicts the hadronization time.}
\end{figure}

Let us set the stage by exploring  the time evolution of the hadronization dynamics in the model. Fig.~\ref{fig:ParticleNumber} depicts the fraction of quark matter on the total number of particles in the system, i.e. quark fraction $=(n_q+n_{\overline q})/(n_{\rm hadron}+n_q+n_{\overline q})$ as a function of time for central Au+Au qMD events at $\sqrt{s_{NN}}=200$~GeV. With the given initial conditions, the fireball stays in a deconfined state during the first 6~fm/c where almost no quarks hadronize. As the system expands further and density decreases, quark recombination into baryons and mesons occurs and the number of deconfined quarks drops to zero. The point at which the parton-hadron transition occurs will be denoted by arrows in the following plots. We now turn to the investigation of the various fluctuation signals.

\subsection{Charged Particle Ratio Fluctuations}

Charged particle ratio fluctuations were proposed as a clear signal for the onset of the quark-gluon plasma phase~\cite{Jeon:2000wg}. The basis for the argument is that the quanta of the electric charge carrier are smaller in a quark gluon plasma phase than in a hadron gas and are distributed over a larger number of particles. Moving one charged particle from/to the rapidity window then leads to larger fluctuations in a hadron gas than in a QGP. The charged particle ratio fluctuations can be quantified by the measure $\tilde{D}$ calculated in the rapidity window $\Delta y$:
\begin{equation}
\tilde{D} = \frac{1}{C_{\mu}C_{y}} \langle N_{ch} \rangle \langle \delta R^2 \rangle_{\Delta y}\quad.
\end{equation}
where $N_{ch}$ stands for the number of charged particles, $R=(1+F)/(1-F)$ with $F=Q/N_{ch}$, $Q$ being the electric charge. Following \cite{Bleicher:2000ek}, the uncorrected charged particle ratio fluctuations $D=\langle N_{ch} \rangle \langle \delta R^2 \rangle$ are divided by the factors $C_{\mu}$ to correct for finite net charge due to baryon stopping and $C_{y}$ for global charge conservation. It was argued that depending on the nature of the initial system, $\tilde{D}$ will yield distinctly different results: $\tilde{D}=1$ for a quark-gluon plasma, $\tilde{D}=2.8$ for a resonance gas and $\tilde{D}=4$ for an uncorrelated pion gas.

Experimentally, charge fluctuations have been measured at RHIC energies by STAR \cite{Pruneau:2003ky,Westfall:2004xy} and PHENIX \cite{Adcox:2002mm,Nystrand:2002pc}.  Both experimental analyses yield results compatible with a hadron gas. Further results  from CERN-SPS \cite{Sako:2004pw,Alt:2004ir} based on a slightly different measure are also compatible with the hadronic expectations.

\begin{figure}
\includegraphics[width=0.5\textwidth]{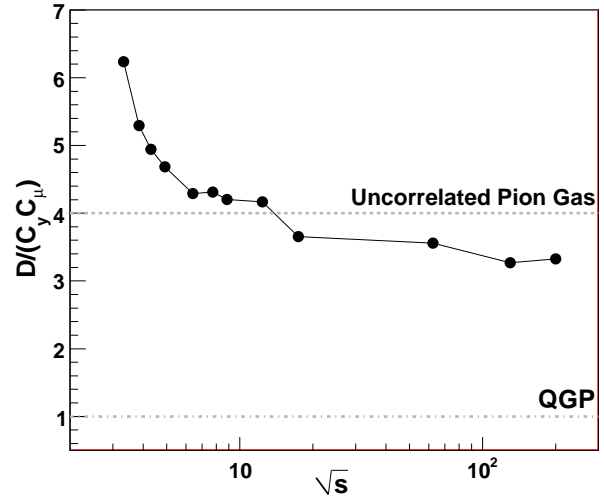}
\caption{\label{fig:Dexcitation} Corrected charged particle ratio fluctuations $\tilde{D}$ as a function of the center of mass energy $\sqrt{s_{NN}}$ within the qMD model for central Au+Au/Pb+Pb reactions (full symbols). Also shown are the values for an uncorrelated pion gas and a quark-gluon plasma.}
\end{figure}

Fig.~\ref{fig:Dexcitation} depicts the energy dependence of the corrected charged particle ratio fluctuations $\tilde{D}$ from AGS energies on ($E_{lab}$=2~AGeV) up to the highest RHIC energy available ($\sqrt{s_{NN}}$=200~GeV) calculated with central Au+Au/Pb+Pb qMD events. As suggested in \cite{Jeon:2000wg,Bleicher:2000ek}, $\tilde{D}$ is calculated in a rapidity window of $y = \pm 0.5$. $\tilde{D}$ decreases from $\tilde{D} \approx 6$ down to $\tilde{D} \approx 3$ with increasing energy and is not compatible with the QGP value in the whole energy range explored, even though the system initially went through a quark phase. No hadronic rescattering stage being included in the present calculations, this is a first indication that the hadronization procedure implemented in the qMD model alone destroys the early stage QGP fluctuations.

\begin{figure}
\includegraphics[width=0.5\textwidth]{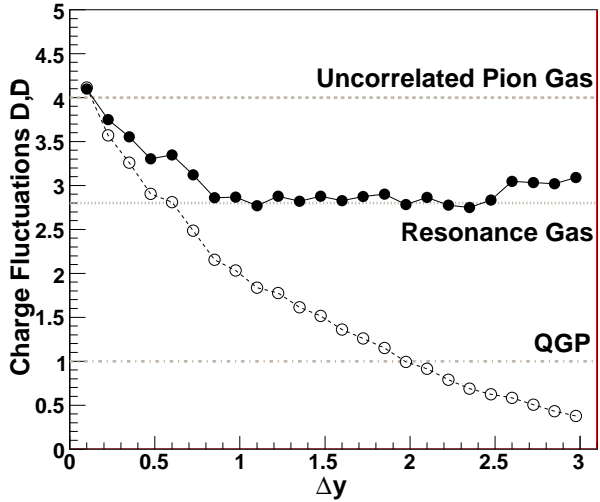}
\caption{\label{fig:RapidityDependence} Corrected $\tilde{D}$ (full symbols) and uncorrected $D$ (empty symbols) charged particle ratio fluctuations as a function of the width of the rapidity window within the qMD model for central Au+Au reactions at $\sqrt{s_{NN}}=200$~GeV. Also shown are the values for an uncorrelated pion gas, a resonance gas and a quark-gluon plasma.}
\end{figure}

The signal might also be sensitive to the size of the rapidity window $\Delta y$ used as can be seen from VNI model calculations \cite{Zhang:2002dy}. The rapidity window dependence for both corrected $\tilde{D}$ (full symbols) and uncorrected $D$ (empty symbols) charged particle ratio fluctuations is plotted on figure Fig. \ref{fig:RapidityDependence} for central Au+Au qMD events at $\sqrt{s_{NN}}=200$~GeV where the quark phase duration is the longest in the present calculations. One can measure the importance of the correction terms as the rapidity window is widened and goes to a regime where global charge conservation becomes so important that $D$ drops to nearly $D \approx 0$. When the rapidity window is wide enough to take into account all particles present in the events, the net charge is fixed by the initial charge of the colliding gold nuclei and can not fluctuate any more. Furthermore, both $\tilde{D}$ and $D$ come very close to 4 when the rapidity window $\Delta y$ is decreased to very small values. If the rapidity window is so small that only one of the decay products of a resonance can be observed, then the system appears completely uncorrelated. In the present model, one observes that $\tilde{D}$ decreases from $\tilde{D}=4$ down to $\tilde{D}\approx 3$ when $\Delta y \approx 1$ and then does not change with increasing rapidity window width until $\Delta y \approx 3$. The result always stays at the expected hadronic expectation value. Thus, the qMD result does not depend on the size of the rapidity window used and is compatible with the hadronic expectations.

\begin{figure}
\includegraphics[width=0.5\textwidth]{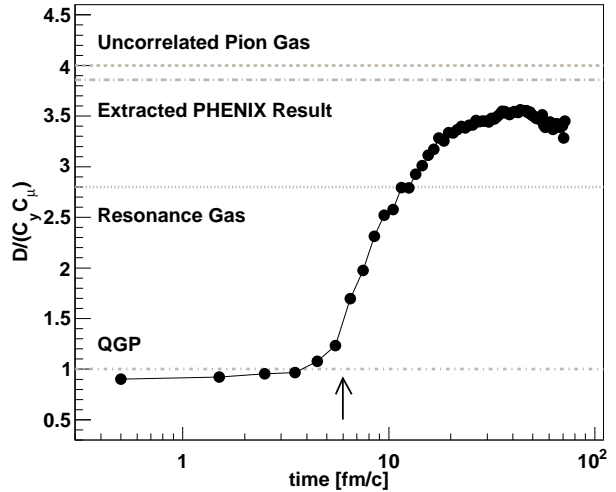}
\caption{\label{fig:Dfluctuations} Corrected charged particle ratio fluctuations $\tilde{D}=D/(C_y C_\mu)$ as a function of time within the qMD model for central Au+Au reactions at $\sqrt{s_{NN}}=200$~GeV (full symbols). Also shown are the values for an uncorrelated pion gas, a resonance gas and a quark-gluon plasma. The arrow depicts the hadronization time.}
\end{figure}

Within the simulations, one has access to the evolution of $\tilde{D}$ from the initial quark phase to the final state. Fig.~\ref{fig:Dfluctuations} shows the result for $\tilde{D}$ from the qMD recombination approach as a function of time calculated for a sample of central Au+Au events at $\sqrt{s_{NN}}=200$~GeV in a rapidity window of $y=\pm 0.5$. In the early stage, when the system is completely in the deconfined phase, $\tilde{D} \approx 1$ as expected. When approaching the hadronization time, $\tilde{D}$ starts to increase and reaches $\tilde{D} \approx 3.5$ after hadronization. The increase of $\tilde{D}$ occurs exactly at the same time as the recombination of the quarks and anti-quarks into hadrons. The slight decrease of $\tilde{D}$ at later times is related to the continuing decay of resonances lowering the charge ratio fluctuations \cite{Jeon:1999gr}.

Gluons are not explicitly included in our simulations and their influence at the transition should be discussed. Following \cite{Jeon:2000wg}, the uncorrected charged particle ratio fluctuations $D$ can be expressed as:
\begin{equation}
D \approx 4 \frac{ \langle \delta Q^2 \rangle }{ \langle N_{ch} \rangle } \quad .
\label{eq:D}
\end{equation}
where $Q$ stands for the net charge and $\langle N_{ch} \rangle$ for the mean number of charged particles. Gluons neither participate in the fluctuations of the charge nor to the number of charged particles in the pure deconfined phase. However, when the system hadronizes, gluons participate in the denominator of Eq. \ref{eq:D} through gluon fragmentation: $g \rightarrow hadrons$, which can lowers $D$. In qMD, the decay of highly excited hadronic clusters, whose masses include the inter-quark potential, effectively models a part of this contribution.

\subsection{Charge Transfer Fluctuations}

As a next observable, we turn to charge transfer fluctuations that were also suggested to provide insight about the formation of a QGP phase. Charge transfer fluctuations are a  measure of the local charge correlation length. They are  defined as \cite{Haussler:2006mq,Jeon:2005yi,Jeon:2005kj,Shi:2005rc}:
\begin{equation}
D_{u}(\eta)= \langle u(\eta)^2 \rangle - \langle u(\eta) \rangle^2\quad,
\end{equation}
with the charge transfer $u(\eta)$ being the forward-backward charge difference:
\begin{equation}
u(\eta)=[Q_F(\eta)-Q_B(\eta)]/2\quad,
\end{equation}
where $Q_F$ and $Q_B$ are the net charges in the forward and backward hemisphere of the region separated at $\eta$ and calculated in the pseudo-rapidity window $ \Delta \eta = \eta \pm$ 1. 

The charge correlation length and hence the local charge fluctuations are expected to be much lower in a quark-gluon plasma than in a hadron gas. Because the measured quantity is local, it can provide information about the presence and the extent of a QGP in (pseudo-)rapidity space. Thus, one expects to observe the lowest value of the charge transfer fluctuations at midrapidity, where the energy density is the highest and where the plasma should be located. Experimental data on this observable are not available up to now.

\begin{figure}
\includegraphics[width=0.5\textwidth]{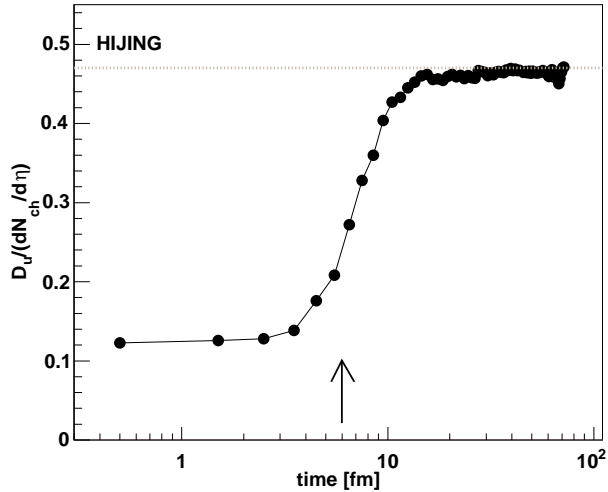}
\caption{\label{fig:ChTransferFluctuations} Charge transfer fluctuations at midrapidity ($\Delta \eta =  \pm 1$) and separation at $\eta=0$ as a function of time within the qMD model (full symbols) for central Au+Au reactions at $\sqrt{s_{NN}}=200$~GeV. The arrow indicates the hadronization time.}
\end{figure}

The time evolution of the charge transfer fluctuations from the present calculations are shown in Fig.~\ref{fig:ChTransferFluctuations}. The sample of events consists again of central Au+Au reactions simulated at $\sqrt{s_{NN}}=200$~GeV. Here, the pseudo-rapidity window corresponds to the STAR acceptance ($ \eta =  \pm 1$). As expected, the correlation length (at central rapidities) is small, with $D_u/(dN_{ch}/dy) \approx 0.1$, as long as the system is in the quark phase. However, similar to the charged particle ratio fluctuations discussed above, the charge transfer measure increases with time up to its hadronic value of $D_u/(dN_{ch}/dy) \approx 0.5$ when hadronization has occured. The final state result is in agreement with the value given by HIJING calculations and therefore in line with the hadronic expectation \cite{Haussler:2006mq,Jeon:2005yi,Jeon:2005kj,Shi:2005rc}.

\begin{figure}
\includegraphics[width=0.5\textwidth]{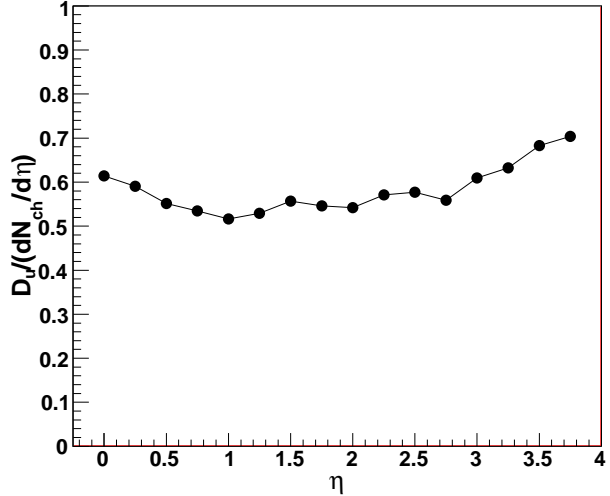}
\caption{\label{fig:ChTransferFluctuationsRapidity}Charge transfer fluctuations for a total rapidity window $\Delta \eta =  \eta \pm 1$ as a function of pseudo-rapidity $\eta$ with the qMD model for central Au+Au reactions at $\sqrt{s_{NN}}=200$~GeV.}
\end{figure}

The dependence of charge transfer fluctuations $D_u/(dN_{ch}/dy)$ on pseudo-rapidity calculated from a sample of central Au+Au qMD events at $\sqrt{s_{NN}}=200$~GeV is depicted on Fig.~\ref{fig:ChTransferFluctuationsRapidity}. The pseudo-rapidity dependence of this observable is flat and compatible with the hadronic expectations ($D_u/(dN_{ch}/dy) \approx 0.5$). In particular, no dip is visible at $\eta \approx 0$ where the plasma is located. Therefore, charge transfer fluctuations seem not only insensitive to the initial QGP distribution in pseudo-rapidity space, but neither to the existence of the quark phase itself in our simulations.

\begin{figure}
\includegraphics[width=0.5\textwidth]{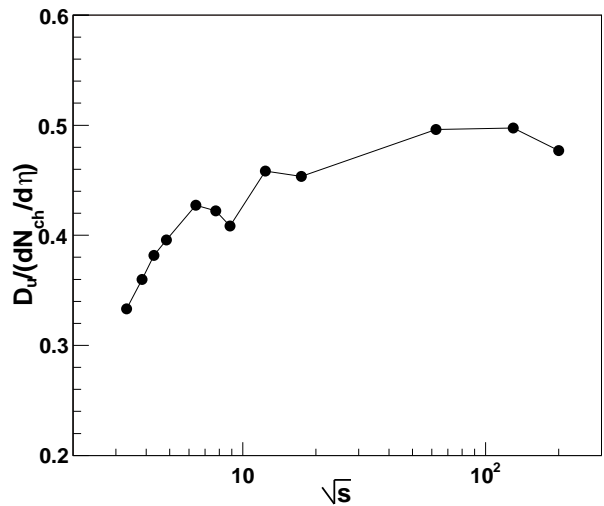}
\caption{\label{fig:ChTexcitation}Charge transfer fluctuations at midrapidity ($\Delta \eta =  \pm 1$) and separation at $\eta=0$ as a function of the center of mass energy $\sqrt{s_{NN}}$  within the qMD model (full symbols) for central Au+Au/Pb+Pb reactions}
\end{figure}

Fig.~\ref{fig:ChTexcitation} depicts the excitation function of $D_u/(dN_{ch}/dy)$ from central Au+Au/Pb+Pb qMD events calculated at midrapidity in the pseudo-rapidity window $\eta =  \pm 1$. $D_u/(dN_{ch}/dy)$ increases from $D_u/(dN_{ch}/dy) \approx 0.3$ at low AGS energies to reach $D_u/(dN_{ch}/dy) \approx 0.5$ at RHIC energies. The result stays above the QGP expectations in the whole energy range explored.
The qMD estimates for charge transfer fluctuations are consistent with the model studied in \cite{Zhou:2006sk}, also including a partonic/hadronic transition.

\subsection{Ratios of Susceptibilities}

We now turn to the study of the baryon number-strangeness correlation coefficient $C_{BS}$, charge-strangeness correlation coefficient $C_{QS}$ and more generally to the ratios of susceptibilities. 

It was argued in \cite{Koch:2005vg} that the correlation between baryon number and strangeness strongly differs whether the system is in a QGP or in a hadronic phase. In a QGP, strange quarks will always carry baryon number $\pm\frac{1}{3}$, which strongly correlates baryon number and strangeness. The situation is different in a hadron gas where strangeness can be carried without baryon number, e.g. with kaons. Following \cite{Koch:2005vg}, we calculate the $C_{BS}$ correlation coefficient on an event-by-event basis:
\begin{equation}
C_{BS}=-3\frac{ \langle B S \rangle - \langle B \rangle \langle S \rangle}{\langle S^2 \rangle - \langle S \rangle^2} \quad ,
\end{equation}
where $B$ stands for the baryon number and $S$ for the total strangeness. The expectation value in an ideal QGP is $C_{BS}=1$ and $C_{BS}=0.66$ in a resonance gas at temperature $T=170$~MeV and chemical potential $\mu=0$ corresponding to the conditions obtained at RHIC \cite{Koch:2005vg}.

Due to difficulties concerning the measure of neutral particles, $C_{BS}$ is difficult to extract from the experimental data. It is however possible to use the related correlation coefficient $C_{QS}$, where $Q$ stands for the electric charge \cite{Gavai:2005yk}:
\begin{equation}
C_{QS}=-3\frac{ \langle Q S \rangle - \langle Q \rangle \langle S \rangle}{\langle S^2 \rangle - \langle S \rangle^2 } \quad ,
\end{equation}

Under the assumption of isospin symmetry \cite{Majumder:2006nq}, $C_{BS}$ and $C_{QS}$ are related through:
\begin{equation}
C_{QS}=\frac{3-C_{BS}}{2} \quad ,
\end{equation}

The variances and covariances of conserved charges can also be expressed in term of susceptibilities and off-diagonal susceptibilities $\chi$: 
\begin{equation}
\begin{array}{rcl}
\langle (\delta X)^2 \rangle & = & V~T~\chi_X \quad , \\
\langle (\delta X) (\delta Y) \rangle & = & V~T~\chi_{XY} \quad , \\
\end{array}
\label{eq:flucssusceptibilities}
\end{equation}
where $V$ is the volume and $T$ the temperature.

Taking the ratio of different susceptibilities then makes the volume term disappear. The $C_{BS}$ and $C_{QS}$ coefficients can thus be written as:
\begin{equation}
C_{BS}= -3 \frac{\chi_{BS}}{\chi_{S}} , \quad C_{QS}= 3 \frac{\chi_{QS}}{\chi_{S}} \quad .
\end{equation}

\begin{figure*}
\includegraphics[width=0.5\textwidth]{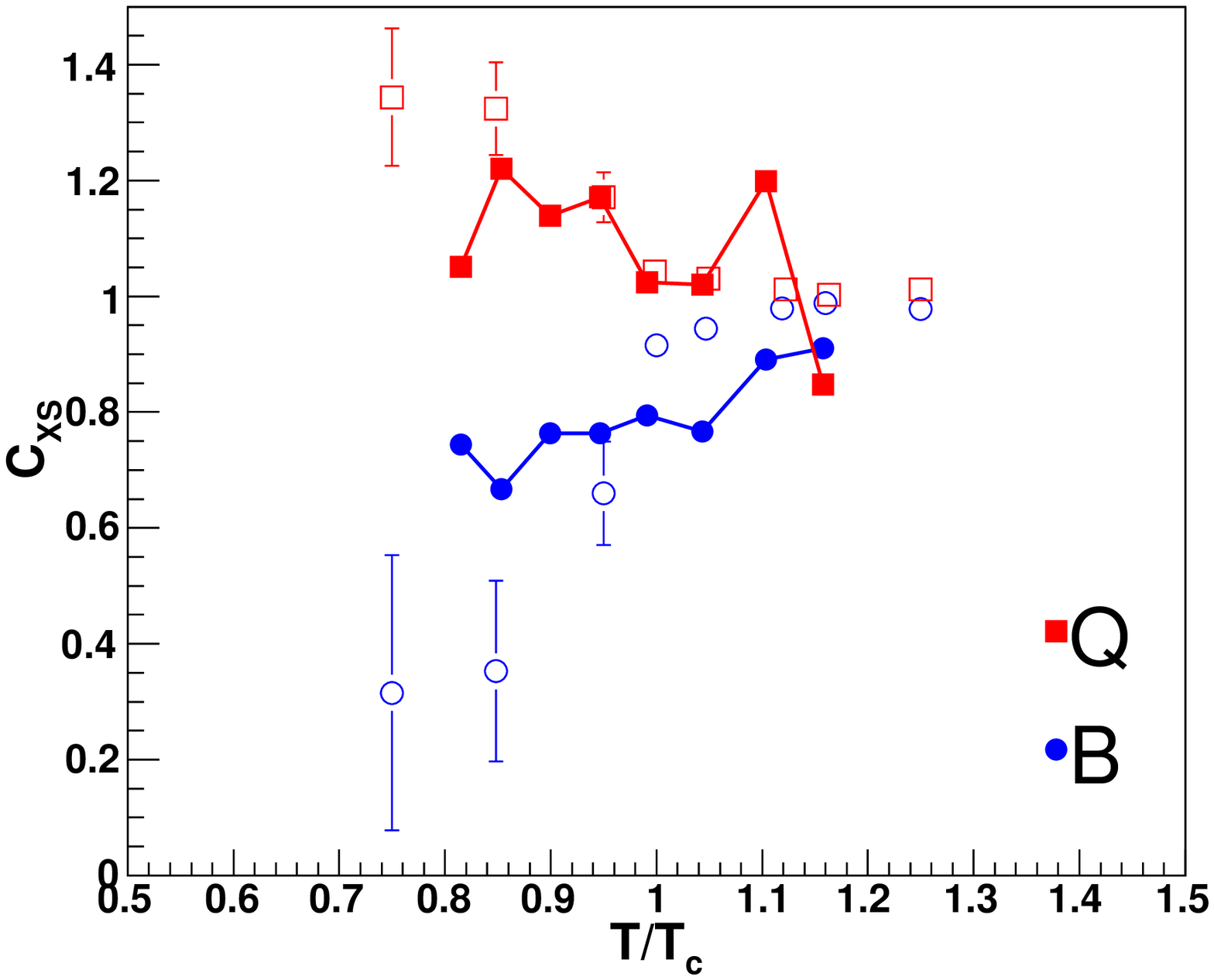}
\includegraphics[width=0.5\textwidth]{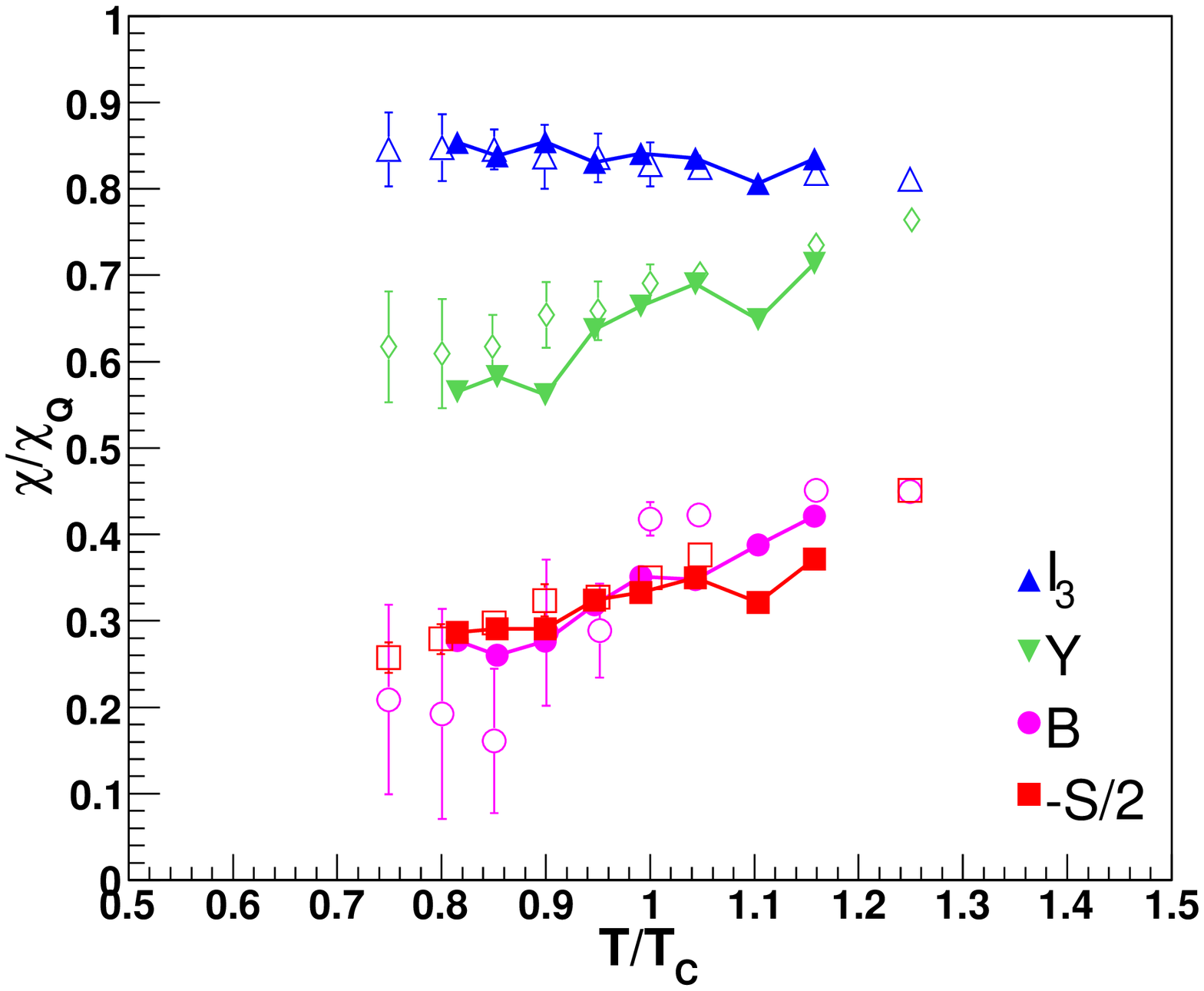}
\caption{\label{fig:CXStemperature} Left: $C_{BS}$ and $C_{QS}$ correlation coefficient as a function of temperature calculated with the qMD model (full symbols) and comparison to lattice data (empty symbols). Right: Various ratios of susceptibilities $\chi_{X}/\chi_Q$ except for $X=S$ which is $(\chi_{S}/\chi_Q)/2$. Open symbols are the result of lattice calculations \cite{Gavai:2005yk}. Full symbols are the result of the qMD model. The temperature is extracted out of the kinetic energy per quark in the medium.
}
\end{figure*}

Fig.~\ref{fig:CXStemperature} (left) depicts $C_{BS}$ and $C_{QS}$ as a function of the temperature as measured by the kinetic energy of the partons in the qMD model (full symbols). In comparison, lattice QCD calculations \cite{Gavai:2005yk} (open symbols) are shown. At high temperature, lQCD yield the results of an ideally weakly coupled QGP with $C_{BS} = C_{QS} = 1$ while below $T_c$, they are consistent within uncertainties with a hadron gas. qMD calculations yield results with $C_{BS} \approx 0.7$ and $C_{QS} \approx 1.2$ in the low temperature regime. Both quantities also go to $C_{BS} \approx C_{QS} \approx 1$ in the quark gluon plasma phase. Similarly to charged particle ratio fluctuations and charge transfer fluctuations, the QGP signal vanishes in the hadronic phase.

Various other ratios of susceptibilities using the electric charge $Q$, strangeness $S$, baryon number $B$, third component of the isospin $I_{3}$ and hypercharge $Y$ have been measured on the lattice and can be used to distinguish between different models, or might serve as signals of the deconfined phase.

Here, we show the result of $\chi_{B}/\chi_{Q}$, $\chi_{I_{3}}/\chi_{Q}$, $\chi_{Y}/\chi_{Q}$ and $-(\chi_{S}/\chi_{Q})/2$ as a function of temperature from qMD calculations (full symbols) and compare with the available lattice data (open symbols) on Fig.~\ref{fig:CXStemperature} (right). $\chi_{I_{3}}/\chi_{Q}$ is essentially flat with $\chi_{I_{3}}/\chi_{Q} \approx 0.9$ with no difference between the deconfined and the confined phase. On the contrary, all other quantities increase from the confined to the deconfined phase with $\chi_{Y}/\chi_{Q}$ going from $\chi_{Y}/\chi_{Q} \approx 0.6$ to $\chi_{Y}/\chi_{Q} \approx 0.7$ in the temperature range explored. $\chi_{B}/\chi_{Q}$ and 
$-(\chi_{S}/\chi_{Q})/2$ are roughly the same and increase from $\chi_{B}/\chi_{Q} \approx -(\chi_{S}/\chi_{Q})/2 \approx 0.2$ to $\chi_{B}/\chi_{Q} \approx -(\chi_{S}/\chi_{Q})/2 \approx 0.4$ with increasing temperature.

Even though the qMD approach is a grossly simplified version of QCD, it compares well with the lattice predictions over the whole temperature range accessible in the course of a collision. This might be seen as an a posteriori justification for the use of the qMD model for the fluctuation and correlation analyses performed above.

\section{Susceptibilities}

Dividing the variances and the covariances of the conserved charges by the average number of charged particles $\langle N_{ch} \rangle$ (see Eq.~\ref{eq:flucssusceptibilities}), which is proportional to $V T^3$, it is possible to access directly information on the diagonal and off-diagonal susceptibilities $\chi$. We define the coefficient $\sigma_{XY}$ by:
\begin{equation}
\sigma_{X Y} = \frac{\langle XY \rangle - \langle X \rangle \langle Y \rangle}{\langle N_{ch} \rangle} \propto \frac{ \chi_{XY}}{T^2}
\label{eq:sigmaxy}
\end{equation}
where $(X,Y)=(B,Q,S,I_{3},Y,\ldots)$.

Note that $\sigma_{QQ}$ corresponds to the net charge fluctuations and is equivalent to the charged particle ratio fluctuations investigated before (see Eq.~\ref{eq:D}).

\begin{figure*}
\includegraphics[width=0.49\textwidth]{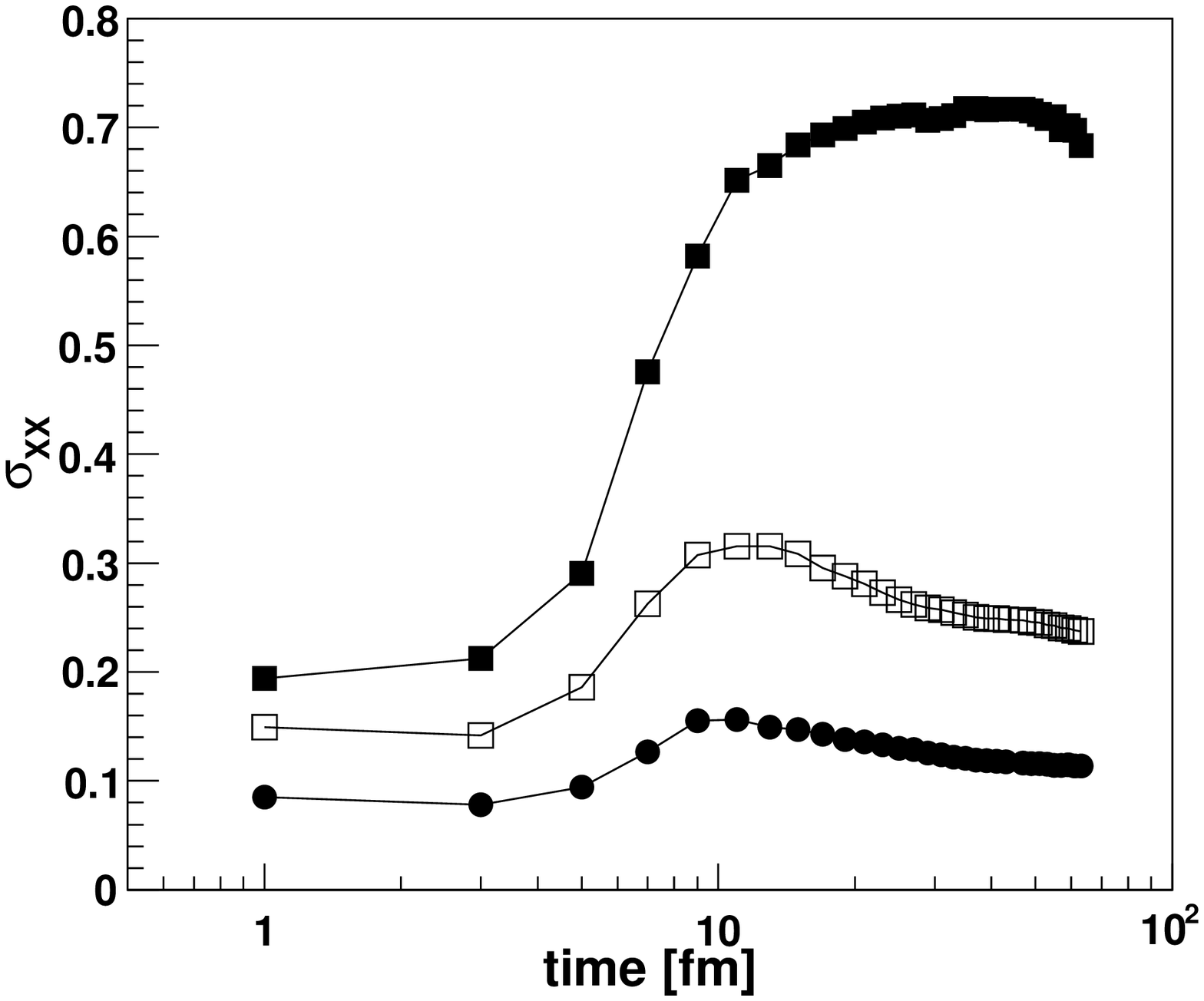}
\includegraphics[width=0.49\textwidth]{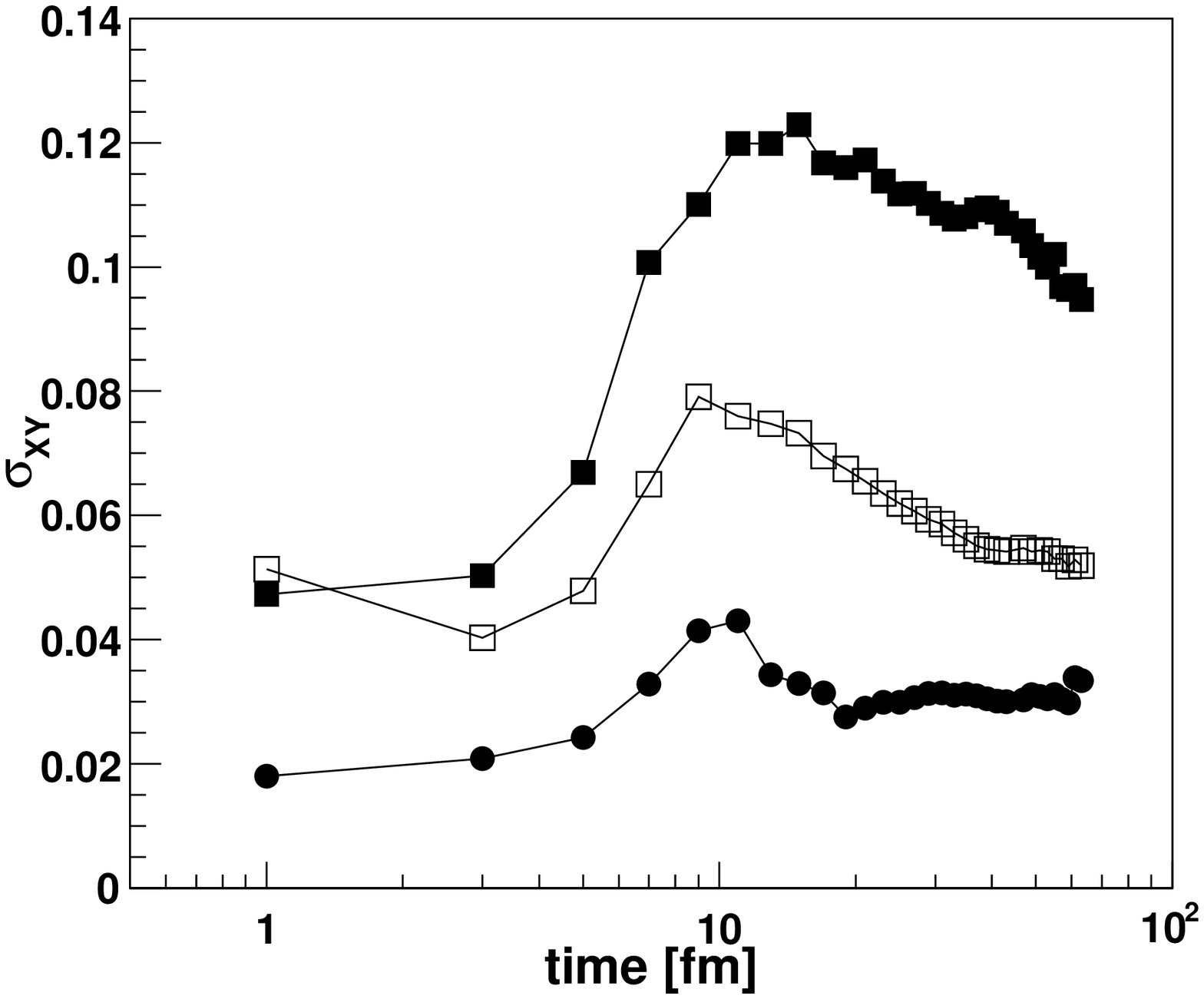}
\caption{\label{fig:susceptibilities} Left: $\sigma_{QQ}$ (full squares), $\sigma_{SS}$ (open squares) and $\sigma_{BB}$ (full circles) as a function of time. Right: $\sigma_{QS}$ (full squares), $\sigma_{BS}$ (open squares) and $\sigma_{QB}$ (full circles) as a function of time. $\sigma_{BS}$ is multiplied by the factor -1. Calculations are performed with a set of central Au+Au qMD events at $\sqrt{s}=200$~AGeV.
}
\end{figure*}

Fig.~\ref{fig:susceptibilities} (left) depicts the time dependence of $\sigma_{QQ}$ (full squares), $\sigma_{SS}$ (open squares) and $\sigma_{BB}$ (full circles). All these quantities exhibit a sharp rise in the vicinity of the hadronization time followed by a smooth decrease to their final hadronic value: $\sigma_{QQ}$ increases from $\sigma_{QQ} \approx 0.2$ to $\sigma_{QQ} \approx 0.7$, consistently with the results obtained for charged particle ratio fluctuations. $\sigma_{BB}$ increases from $\sigma_{BB} \approx 0.08$ to $\sigma_{BB} \approx 0.12$ and $\sigma_{SS}$ from $\sigma_{BB} \approx 0.14$ to $\sigma_{BB} \approx 0.24$. Besides the notable exception of $\sigma_{QQ}$, they all exhibit a characteristic enhancement at the transition from partonic to hadronic matter.

The result of the correlation coefficients $\sigma_{XY}$ is presented in Fig.~\ref{fig:susceptibilities} (right) for $\sigma_{QS}$ (full squares), $\sigma_{BS}$ (open squares) and $\sigma_{QB}$ (full circles). $\sigma_{BS}$ is multiplied by the factor -1. The behaviour is qualitatively similar with $\sigma_{QB}$ increasing from $\sigma_{QB} \approx 0.02$ in the pure deconfined phase up to $\sigma_{QB} \approx 0.03$ in the hadronic stage. $\sigma_{QS}$ increases with time from $\sigma_{QS} \approx 0.05$ up to $\sigma_{QS} \approx 0.1$. $\sigma_{BS} \approx 0.05$ in both partonic and hadronic phase but still exhibit an enhancement around the transition time. The situation for $\sigma_{XY}$ strongly contrasts with the ratios of susceptibilities which smoothly increase from their QGP to their hadronic values.

These correlation and fluctuation studies of $\sigma_{XY}$ can be extended to isospin $I_3$ and hypercharge $Y$ from which experimentally measurable quantities can be constructed, e.g. the $C_{MS}$ correlation coefficient \cite{Majumder:2006nq}. However, the investigation of $B$, $Q$ and $S$ is sufficient to treat the problem completely for three light flavors of quarks $u$ ,$d$ and $s$. In the QGP stage, $Q$, $B$ and $S$ can be rewritten in terms of net upness $\Delta u$, net downness $\Delta d$ and net strangeness $\Delta s$ \cite{Gavai:2005yk} whose fluctuations and correlations translate into quark susceptibilities. Note that equivalent sets of conserved charges can be chosen, e.g. ($B, I_3, Y$) and extended to heavy quark flavours.

\section{Vanishing of the signals}

Finally, we discuss the transition from the initial QGP fluctuations and correlations to their hadronic values for the observables calculated here. We first investigate the effect of the decreasing mean number of charged particles $\langle N_{ch} \rangle$ at hadronization entering the definition of charged particle ratio fluctuations, charge transfer fluctuations and $\sigma_{XY}$ coefficients.

\begin{figure}
\includegraphics[width=0.5\textwidth]{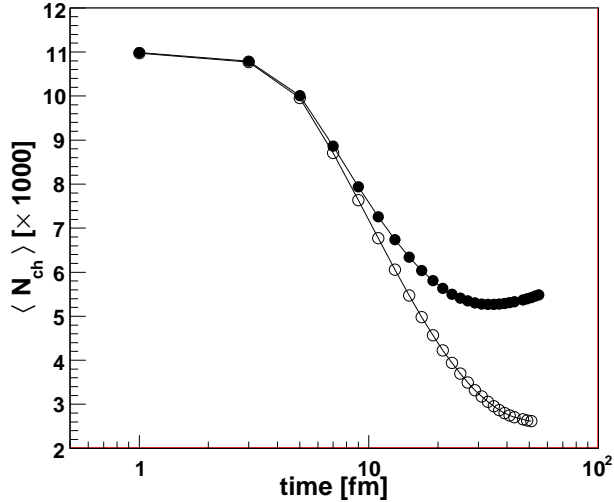}
\caption{\label{fig:NchEvolution} Mean number of charged particles $\langle N_{ch} \rangle$ as a function of time. Calculations are performed with a set of central Au+Au qMD events at $\sqrt{s}=200$~AGeV with resonance decay included (full circles) and without (open circles).}
\end{figure}

To pin down the effect of the dynamical recombination process itself, free from the influence of resonance decay, we use a set of central Au+Au qMD events at $\sqrt{s_{NN}}=200$~GeV where the hadronic clusters formed are not allowed to decay. The time dependence of $\langle N_{ch} \rangle$ is depicted on Fig.~\ref{fig:NchEvolution} with resonance decay (full circles) and without resonance decay (open circles). In the absence of decaying hadronic clusters, the number of particles decreases from $\langle N_{ch} \rangle \approx 11000$ down to $\langle N_{ch} \rangle \approx 2700$, whereas it decreases from $\langle N_{ch} \rangle \approx 11000$ down to $\langle N_{ch} \rangle \approx 5500$ for the default calculations. The initial $N_q$ charged quarks recombine into $N_q/2$ hadrons, among which half, i.e. $N_q/4$, are charged. The variances and covariances of the conserved charges exhibit no sharp structure at the partonic/hadronic transition. Thus, the decrease of the number of charged particles at hadronization is  responsible for the sharp jump of the $\sigma_{XY}$ coefficients, charged particle ratio fluctuations $\tilde{D}$ and charge transfer fluctuations $D_u/(dN_{ch}/dy)$. For full qMD calculations, resonance decay affects both the numerators and denominator of the studied observables and slightly lowers down the result to values which are still compatible with the hadronic value.

For the ratios of susceptibilities such as $C_{BS}$, this argument is not valid as $\langle N_{ch} \rangle$ plays no role. The transition from the QGP to the hadronic values can result from correlations induced by the interaction potential itself, i.e. the gathering of quarks into color neutral clusters, the shuffling of the quarks in and out of the considered rapidity window at hadronization, or be affected by resonance decay. To distinguish between these different mechanisms, we use sets of central Au+Au qMD events at $\sqrt{s_{NN}}=200$~GeV with no mapping to hadronic clusters --the quark system is let to evolve according to the qMD Hamiltonian of Eq.~\ref{eq:hamiltonian}-- to set apart the effect of the interaction potential, with mapping and no resonance decay to investigate the rapidity shift of the quarks from their initial value to the hadronic clusters rapidity, and with full calculations including resonance decay.

\begin{figure}
\includegraphics[width=0.49\textwidth]{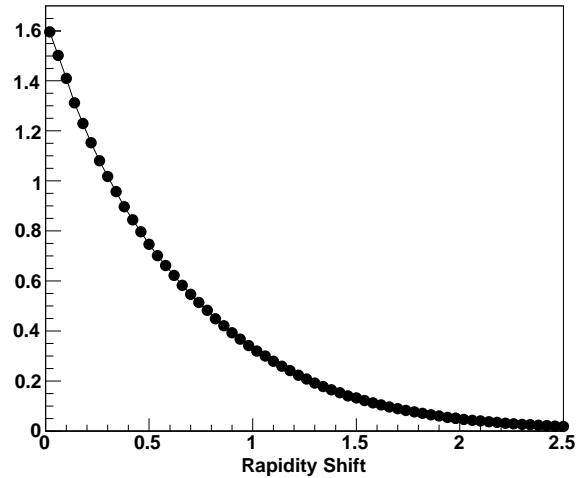}
\caption{\label{fig:RapidityShift} Distribution of the rapidity shift of the quarks at hadronization $y_{quark}-Y_{cluster}$. The mean rapidity shift is $\langle | y_{quark} - Y_{cluster}| \rangle = 0.57$.}
\end{figure}

Let us discuss the capability of the qMD model to move charges in and out of the studied rapidity window at hadronization. Fig~\ref{fig:RapidityShift} depicts the distribution of the rapidity shift of the quarks at hadronization $y_{quark}-Y_{cluster}$ obtained in the qMD model. The mean rapidity shift is $\langle | y_{quark} - Y_{cluster}| \rangle = 0.57$ and can partly explain the vanishing of the signals.

\begin{figure*}
\flushleft{\includegraphics[width=0.49\textwidth]{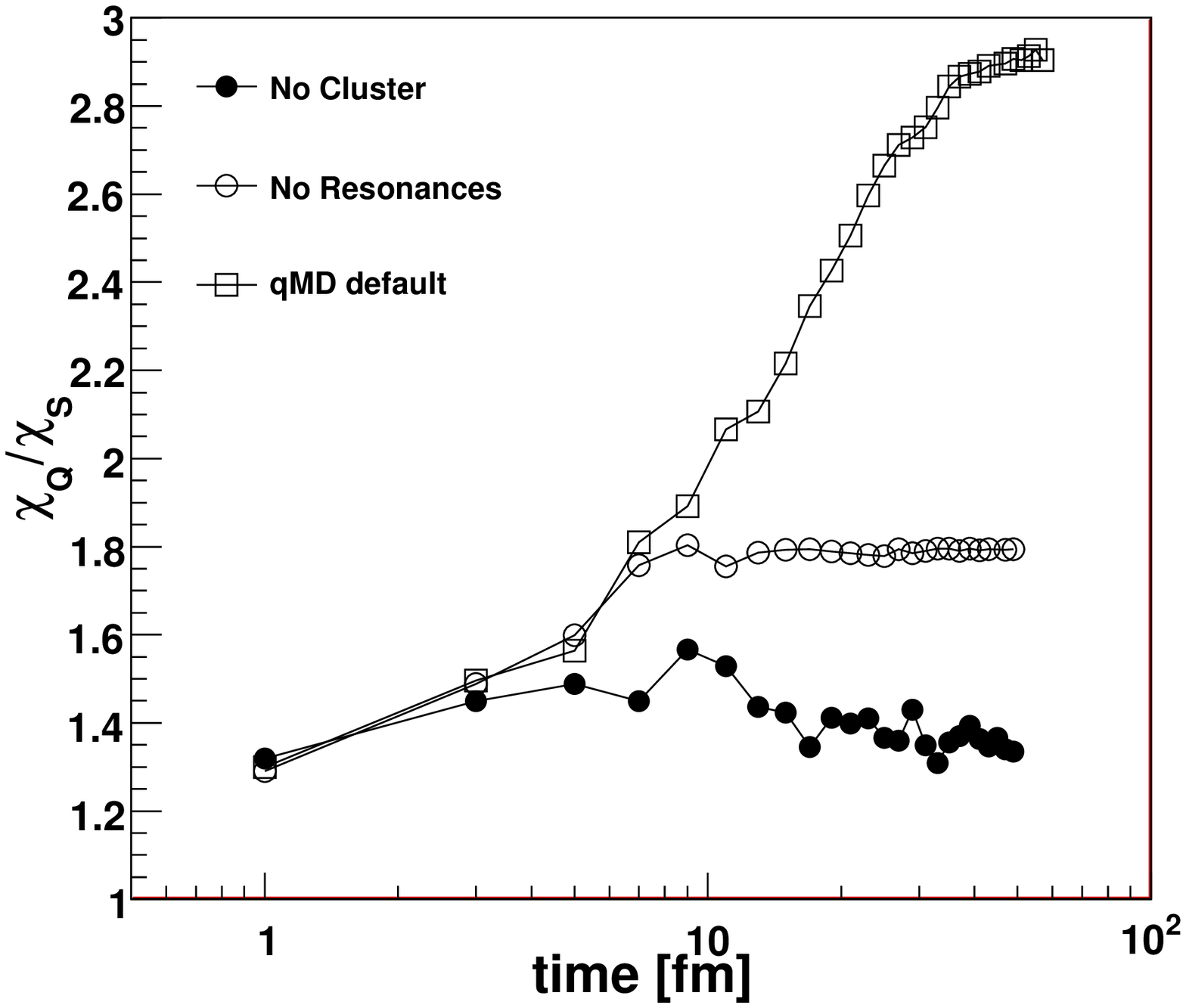}}
\includegraphics[width=0.49\textwidth]{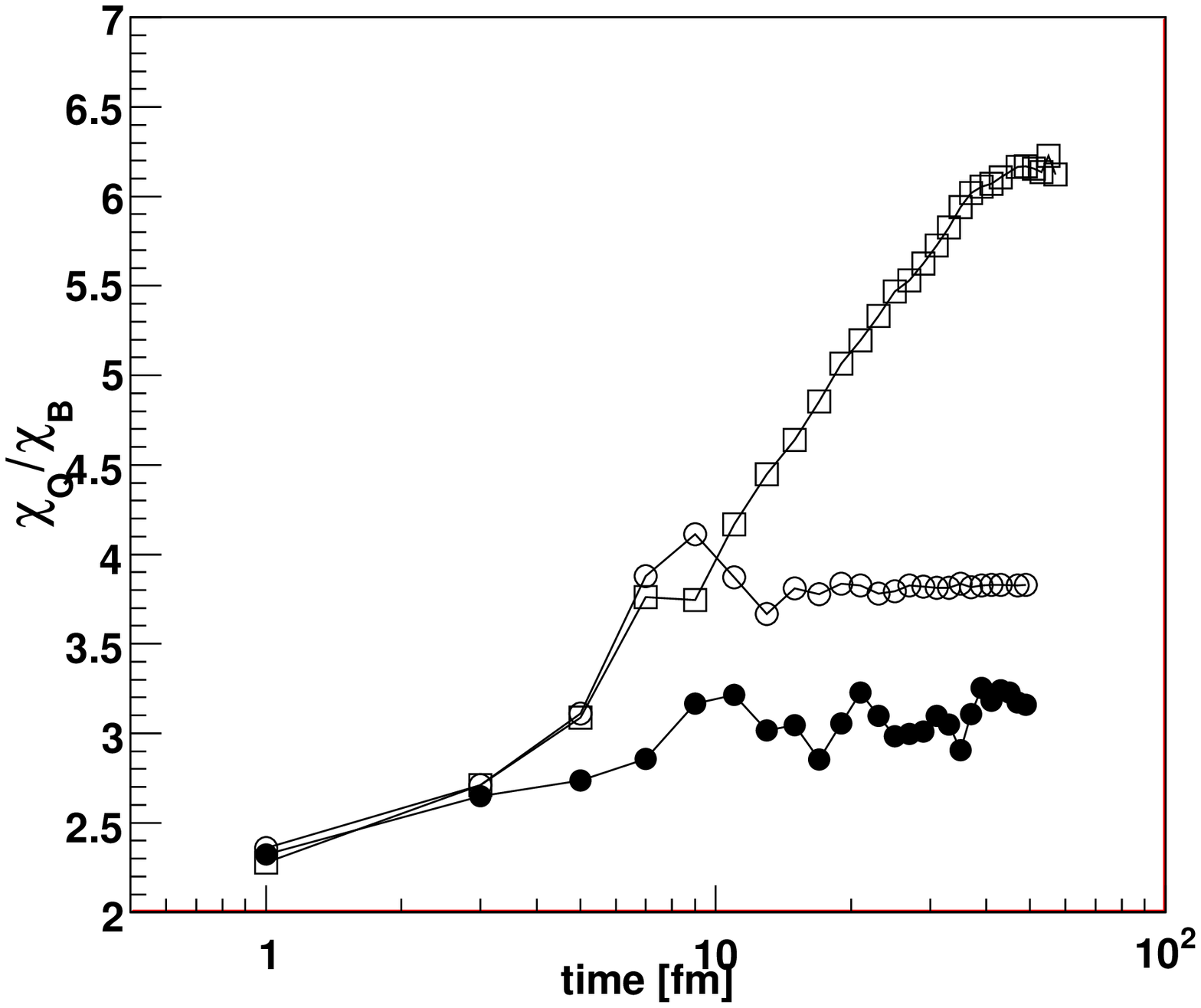}
\flushleft{\includegraphics[width=0.49\textwidth]{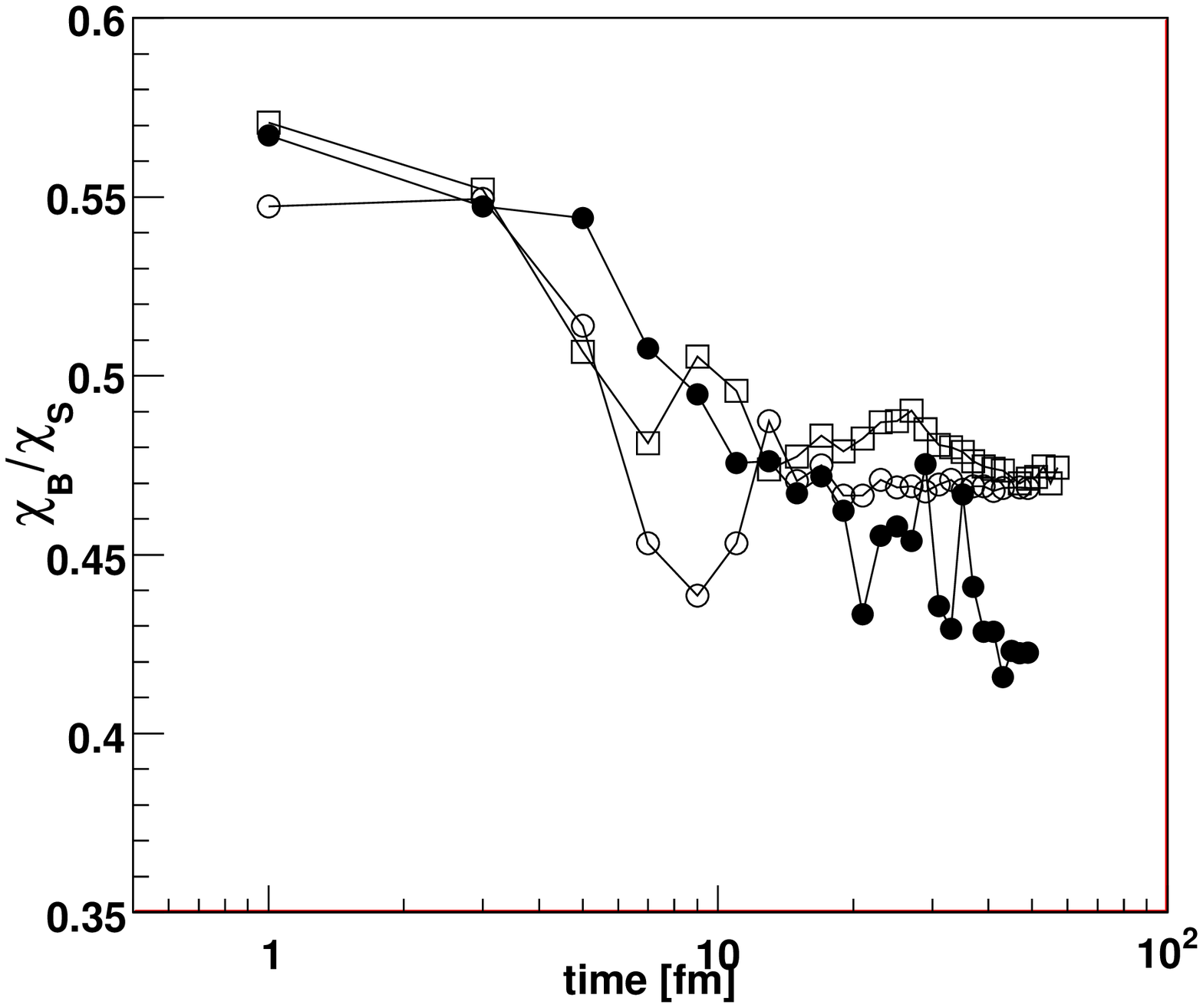}}
\caption{\label{fig:XoY} Top Left: $\chi_Q/\chi_S$ as a function of time Top right: $\chi_Q/\chi_B$  as a function of time. Bottom: $\chi_B/\chi_S$ as a function of time. Calculations are performed with sets of central Au+Au qMD events at $\sqrt{s}=200$~AGeV without hadronization (full circles), with mapping and no resonance decay (open circles) and for full qMD calculations (open squares).
}
\end{figure*}

Figs~\ref{fig:XoY} shows the time evolution of $\chi_Q/\chi_S$ (top left), $\chi_Q/\chi_B$ (top right) and $\chi_B/\chi_S$ (bottom) from qMD calculations where the quark system is let to evolve according to Eq.~\ref{eq:hamiltonian} (full circles), when the quark cluster are mapped to hadronic states but resonances not allowed to decay (open squares) and finally for full qMD calculations with mapping and resonance decay included (open circles).

The effect of the rapidity shift depicted in Fig.~\ref{fig:RapidityShift} is the difference between the calculations without mapping (full circles) and without resonance decay (open circles). $\chi_Q/\chi_S$ without mapping stays flat around $\chi_Q/\chi_S \approx 1.3$ whereas $\chi_Q/\chi_S$ increases with time from $\chi_Q/\chi_S \approx 1.3$ up to $\chi_Q/\chi_S \approx 1.8$. The further increase to $\chi_Q/\chi_S \approx 2.9$ in full qMD calculations is related to the decay of resonances. $\chi_Q/\chi_B$ increases smoothly from $\chi_Q/\chi_B \approx 2.3$ up to $\chi_Q/\chi_B \approx 3$ without mapping and from $\chi_Q/\chi_B \approx 2.3$ up to $\chi_Q/\chi_B \approx 3.8$ with mapping. The further increase to $\chi_Q/\chi_B \approx 6$ is due to the continuing decay of resonances. The situation is markedly different for $\chi_B/\chi_S$ which decreases with time from $\chi_B/\chi_S \approx 0.57$ down to $\chi_B/\chi_S \approx 0.47$ with no visible difference between all sets of events. In this last case, the decrease of $\chi_B/\chi_S$ with time is related to the potential-driven gathering of the quarks into color neutral clusters. We thus conclude that the ratios of susceptibilities is influenced by these three different effects with varying importance. the correlation due to the inter-quark potential used, the rapidity shift of the quarks to the rapidity of the hadronic cluster at hadronization, and the decay of resonances.

\section{Conclusions}

We have investigated various quantities constructed from the fluctuations and correlations of the conserved charges within the dynamical recombination model quark Molecular Dynamics  (qMD) which includes an explicit microscopic transition from partonic to hadronic matter. The sample of events used consists of central Au+Au/Pb+Pb reactions simulated with the qMD model from low AGS energies on ($E_{lab}=2$~AGeV) up to the highest RHIC energy available ($\sqrt{s_{NN}}=200$~GeV).

We first discussed our approach and addressed the problem of entropy conservation at the partonic/hadronic transition with quark recombination. We find that the hadronization procedure implemented in the qMD model is compatible with the necessary increase of entropy with time.

The charged particle ratio fluctuations $\tilde{D}$, the charge transfer fluctuations $D_u/(dN_{ch}/dy)$, the baryon number-strangeness correlation coefficient $C_{BS}$,  the charge-strangeness correlation coefficient $C_{QS}$ were then computed and found to be compatible with their hadronic expectation values in the qMD final state, even though in line with the expected QGP results in the early stage. The vanishing of the initial QGP correlations and fluctuations occurs at the partonic/hadronic transition and was traced back to the recombination-hadronization process. Our results are supported by the agreement of our model with various ratios of susceptibilities calculated in lattice QCD.

The investigation of the variances and covariances divided by the number of charged particles $\sigma_{XY} = (\langle X Y \rangle - \langle X \rangle \langle Y \rangle)/\langle N_{ch} \rangle$ permits to directly access information on diagonal and off-diagonal susceptibilities. We show that these quantities exhibit a sharp increase from their QGP to their hadronic value at the transition. All the susceptibilities present a maximum in the vicinity of the hadronization time with the notable exception of the electric charge susceptibility $(\langle Q^2 \rangle - \langle Q \rangle^2)/\langle N_{ch} \rangle$. We stress that the study of these quantities,  only taking into account the electric charge $Q$, baryon number $B$ and strangeness $S$ (or another equivalent set of conserved charges), allows to treat completely the problem of initial QGP correlations and fluctuations for a system of light quark flavours $u$, $d$ and $s$. This work can be extended to heavy quark flavours.

The transition from the initial QGP to the expected hadronic correlation and fluctuation values results from various effects. The vanishing QGP signal for charged particle ratio fluctuations, charge transfer fluctuations as well as for the $\sigma_{XY}$ coefficients are driven by the decrease of the number of charged particles $\langle N_{ch} \rangle$ at hadronization. For the ratios of susceptibilities, independent of $\langle N_{ch} \rangle$, the vanishing signal is related to the gathering of quarks into color neutral clusters due to the interaction potential used, the rapidity shift of the quark to the hadronic cluster rapidity at hadronization, and the decay of resonances. The relative importance of these effects differs for different ratios.

The fact that no similar fluctuation signals was ever observed in the experimental data until now can thus be understood as the result of (recombination-)hadronization. It would be interesting to see the experimental result of the fluctuation and correlation quantities investigated here as they are complementary to the charged particle ratio fluctuations, already extensively measured.

\begin{acknowledgments}
We thank Prof. S. Pratt, Dr. G. Torrieri, Dr. Molnar and M. Hauer for fruitful discussions and comments. The computational resources have been provided by the Center for Scientific Computing (CSC) at frankfurt. This work was supported by GSI and BMBF.
\end{acknowledgments}

\end{document}